# Manifold learning-based feature extraction for structural defect reconstruction

Qi Li[1], Dianzi Liu[2,*], Zhenghua Qian[1,*]

[1]State Key Laboratory of Mechanics and Control of Mechanical Structures, College of Aerospace Engineering,

Nanjing University of Aeronautics and Astronautics, Nanjing, 210016, China

[2]School of Engineering, University of East Anglia, UK

* Corresponding authors: dianzi.liu@uea.ac.uk and qianzh@nuaa.edu.cn

**Abstract:** Data-driven quantitative defect reconstructions using ultrasonic guided waves has recently demonstrated great potential in the area of non-destructive testing. In this paper, we develop an efficient deep learning-based defect reconstruction framework, called NetInv, which recasts the inverse guided wave scattering problem as a data-driven supervised learning progress that realizes a mapping between reflection coefficients in wavenumber domain and defect profiles in the spatial domain. The superiorities of the proposed NetInv over conventional reconstruction methods for defect reconstruction have been demonstrated by several examples. Results show that NetInv has the ability to achieve the higher quality of defect profiles with remarkable efficiency and provides valuable insight into the development of effective data driven structural health monitoring and defect reconstruction using machine learning.

## 1. Introduction

The inverse scattering problem is determining the characteristics of a media (flaw shape, material properties, etc.) from scattered wave data. Mathematically, the inverse scattering problem lies in the frame of inverse problem, which has been systematically described and reviewed by Tarantola [1] and Valette[2]. Using ultrasonic guided wave to detection and realize the reconstruction of structural defect is an inverse scattering problem. Currently, one of the researches on guided wave inverse scattering is attributed to Rose and Wang[3]. They deduced a dyadic Green's function for a point force in a plate using Mindlin plate model, by the help of which they derived the far-field radiation pattern due to piezoelectric radiators. Applying the Green's function and the Born approximation, they showed a relationship between far-field scattering amplitude and spatial Fourier transform of local homogeneity, based on which an approach to reconstruct weak flexural inhomogeneity is proposed[4]. Wang et al.[5] used the Born approximation to replace the total field near the defect which is regarded as a weak scattering source with the incident field, derived the mathematical relationship between the reflection coefficient located in the far field and the defect shape function into Fourier transform pairs, and reconstructed the thinning defect in the two-dimensional plate. In the field of pipeline defect detection, Da [6] et al. proposed a novel method (QDFT) for the quantitative reconstruction of pipeline defects based on ultrasonic guided SH-waves. This method started from the boundary integral equation and the Fourier transform of wavenumber domain, and derived the defect shape function using the Born approximation. Finally, the unknown defect was reconstructed through the reference model.

In addition to adopting theoretical analysis to solve inverse scattering problems, some machine learning based approaches have also been researched. David et al. [7] studied the combination of FBP algorithm and PWLS iterative algorithm with convolutional neural network to reconstruct images. The study found that the local fusion between this algorithm can improve the balance between resolution and variance in the image reconstruction process, so it can improve the quality of the CT image. In order to solve the ill-posed problem in the computational tomography, Jin et al. [8] combined the deep convolutional neural network with the filtered back projection algorithm (FBP). First use the back projection algorithm to process the sub-sampled sinogram to obtain a preliminary reconstructed image, and then input the reconstructed image into the convolutional neural network for post-processing, and output a high-quality reconstructed image. In the review article of Michael [9], a lot of research work using deep learning algorithms for inverse scattering problem is listed, and it is explained that in the field of image reconstruction, the mainstream learning based method is to combine the traditional reconstruction algorithm with the deep learning algorithm. First use traditional theoretical methods for pre-reconstruction, and then input the reconstruction results into the trained machine learning model for post-processing to obtain high-quality reconstruction results.

Different from the past method, here a novel deep convolutional neural network framework is proposed to directly realize the mapping from the data of ultrasonic guided wave scattering field to the shape function of the local defect, and realize the high accuracy and high efficiency defect reconstruction.

## 2. Forward scattering analysis

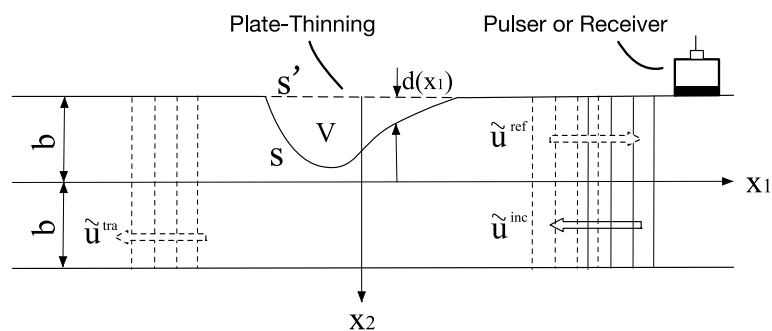

Fig.1 Reflection and transmission of an incident guided SH-wave by a plate thinning.

The problem configuration is shown in Fig.1. The ultrasonic guided SH-waves is excited on the right side of the plate and reflected by the thinning[5]. Starting from the wave equation and corresponding boundary condition in the plate:

$$\left(\frac{\partial^2}{\partial x_1^2} + \frac{\partial^2}{\partial x_2^2}\right)u - \frac{1}{V_s^2}\frac{\partial^2 u}{\partial t^2} = 0 \tag{1a}$$

$$\tau_{23} = \frac{\mu \partial u}{\partial x_2} = 0 \text{ at } x_2 = \pm b \tag{1b}$$

Here $V_S$ is the SH-wave velocity, $2b$ is the plate depth, the displacement field of $n$th SH-wave mode is obtained in the form of

$$\tilde{u}_n = A_n f_n(\beta_n x_2) e^{\pm i \xi_n x_1} \tag{2a}$$

Where $\beta_n = n\pi/2b$, $\xi_n = \sqrt{\frac{\omega^2}{V_s^2} - \beta_n^2}$, and $A_n$ is the amplitude coefficient. $f_n(x)$ is a function defined as:

$$f_n(x) = \begin{cases} \cos x & \text{for } n = 0,2,4,\dots \\ \sin x & \text{for } n = 1,,3,5,\dots \end{cases} \tag{2b}$$

Assuming that the incident guided SH-waves in this problem is a simple mode propagating from right to left, the displacement field of the incident and reflected waves can be expressed as:

$$\tilde{u}^{\text{inc}} = A_n^{\text{inc}} f_n(\beta_n x_2) e^{+i\xi_n x_1} \qquad \tilde{u}^{\text{ref}} = A_n^{\text{ref}} f_n(\beta_n x_2) e^{-i\xi_n x_1} \tag{3}$$

The reflection coefficient is defined as the ratio of the two coefficients:

$$C^{\text{ref}} = A_n^{\text{ref}}/A_n^{\text{inc}} \tag{4}$$

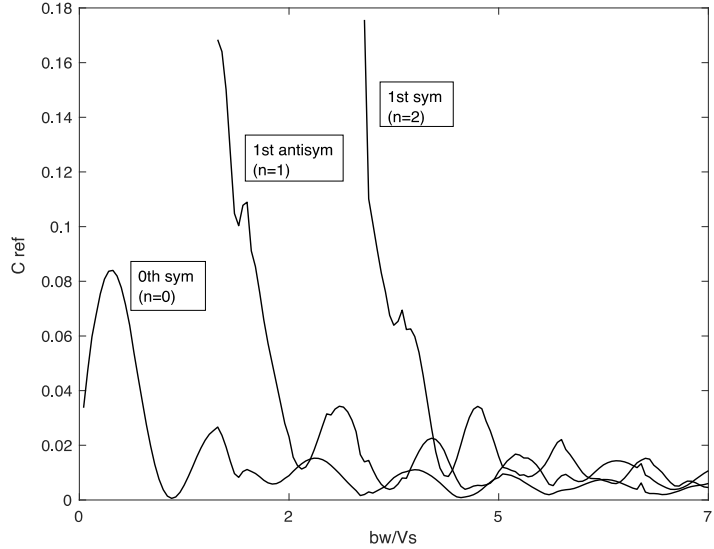

Fig.2 Reflection coefficients of the first five SH-wave modes due the plate thinning.

We use the mode exciting method to obtain the amplitudes $A_n^{\text{ref}}$ of reflected waves as a function of wavenumber $\xi_n$. In this method, a truncated waveguide with enough length is chosen as a model for calculation, in which a thinning area is located right at the middle of the model. By exerting a proper displacement set on both the truncated ends, the wave filed within the truncated waveguide is obtained ends, the wave field within the truncated waveguide is obtained by hybrid finite element method (HFEM). As shown in Fig.2, the absolute values of reflection coefficients for the first five modes, namely, $0^{th}$ symmetric(n=0), 1st anti-symmetric(n=1),1st symmetric(n=2), when the scatterer is a V-shaped thinning.

## 3. Deep learning-driven inverse scattering model

Guided wave defect reconstruction can be attributed to a scattering problem. For the scattering problem, it can be simply expressed by the following formula:

$$y = Tx + \xi \tag{5}$$

Where $x$ represents the scattering source, in this paper, it represents the thinning defect in the board; $y$ represents the scattering field signal; $T$ is an operator, and the properties of the operator $T$ depend on the specific scattering problem; $\xi$ is the error. The task of scattering inversion is to calculate $x$ based on $y$. The traditional ways to solve this problem are divided into two types. The first way is to solve directly, that is, to construct the inverse problem model directly. The expression of such methods is:

$$x = \hat{T}^{-1} y \tag{6}$$

Where $\hat{T}^{-1}$ is the theoretical reconstruction model. The advantage of this method is that for the reconstruction of defects in simple structures, the calculation of the scattering inversion can be completed in a short time; The disadvantage is that scattering inversion is an ill-posed problem, it is difficult to calculate accurate results. In particular, when the scattering problem becomes complex, it will be extremely difficult to construct the reconstructed model, and the reconstruction accuracy will be further affected.

The second type of traditional method for solving the scattering inversion problem is an iterative-based method, such as the QDFT [9], which is expressed as:

$$O\{y\} = \arg\min_{x} f(T\{x\}, y) \tag{7}$$

The function $f$ in the formula is used to characterize the error between $T\{x\}$ and $y$. The advantage of the iterative-based method is that it can obtain accurate results; the disadvantage is that the iterative process requires a lot of calculations, and the time cost is higher.

In this paper, we propose an inverse scattering model for guided wave defect reconstruction based on deep learning framework (NetInv), which can be expressed as:

$$L = \arg\min_{\theta} \sum_{n=1}^{N} f(x_n, L_\theta\{y_n\}) + g(\theta) \tag{8}$$

Where $x_n$ represents the shape of plate thinning, $y_n$ represents the reflection coefficient

$C^{ref}$. $N$ represents the number of samples; $L_\theta$ is the neural network built for inversion calculations, $\theta$ is the parameter in the neural network, and it is the iterative update object during the training process; $f$ is the error function, used to characterize the difference between samples $x_n$ and $L_\theta\{y_n\}$; $g$ is a regularization term, which limits the value of parameter $\theta$ to reduce the complexity of the trained model $L_\theta$ and prevent over fitting. Fig.3 presents the process of defect reconstruction using NetInv.

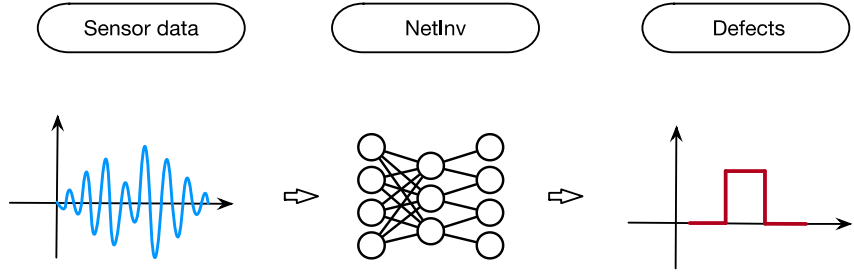

Fig.3 Manifold learning-assisted defect reconstruction using NetInv.

## 4. Experimental validation

We now present the result of validating our method on simulated datasets. The simulated dataset consists of V-shaped defects, Gaussian-curved defects, Rectangular-shaped defects and their reflection coefficients respectively.

In order to quantify the difference between the reconstructed defect and the real defect, that is, the quality of the reconstructed defect, the signal-to-noise ratio (SNR) [27] is defined:

$$SNR(x,\hat{x}) \triangleq \max_{a \in \mathbb{R}} \left\{10 log_{10}(\frac{\|x\|_{l_2}^2}{\|x - a\hat{x}\|_{l_2}^2})\right\} \qquad (9)$$

Where $x$ is the real defect, and $\hat{x}$ is the reconstructed defect. A higher SNR value corresponds to a better reconstruction. Note that the vector $x$ or $\hat{x}$ used to characterize the defect shape in this study is actually the spatial distribution of the defect shape in the entire detection range, including the defect area and the defect-free area. The purpose of this design is to not only study the quality of reconstruction defects in the defect area, but also study the noise and error in the non-defect area of the reconstruction result.

### 4.1 Reconstruction results

The datasets with three types of defects are obtained by finite element method. It contains 800 defects, separated into 700 defects for training, 60 defects for validation, and 40 defects for testing. The depth of defects is range from 0 to 0.8(The depth of plate is 1). We trained a NetInv with the reconstruction mean squared error (MSE) as the loss function and using the value of SNR to quantitively measuring the quality of the reconstructed defects.

We evaluated the proposed model by comparing the results with a conventional reconstruction model – wavenumber spatial transform method based on Born approximation (WNST). Figure 4 presents the comparison between the actual flaw shapes and the reconstructed results with two methods for three types of rectangular-shaped, Gaussian-curved and V-shaped flaws with the different width and different maximum depths. We can see the locations and the basic features of flaws are found in all cases with two methods, but the results from NetInv are closer to actual flaws. Table 1 summarizes the SNR of reconstructed defects corresponding to each method. The results show that the output of NetInv substantially outperforms the baseline method and leads to higher SNR values. Especially, for the rectangular-shaped defect and Gaussian-curved defect, the results form NetInv are nearly $200\%$ more accurate than the solution by WNST. In addition, the computational cost of NetInv is extremely low during the reconstruction stage, where each reconstruction corresponds to a simple forward pass through the CNN. In our case, the reconstructing time of NetInv for a single defect is less than 1 second.

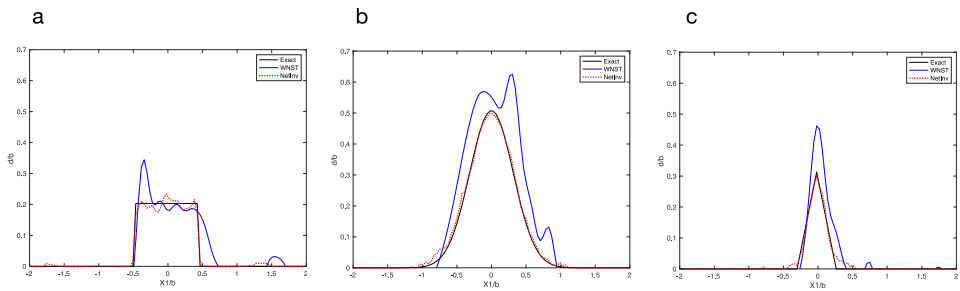

Fig.4 Comparison of the reconstructed defects using the reconstruction approach WNST and the NetInv.

Table 1 Comparison of SNR (dB) values of reconstruction results of the two methods

| Reconstruction Methods | Rectangular defects | Gaussian–curved defects | V–shaped defects |
| --- | --- | --- | --- |
| WNST | 6.82 | 9.65 | 10.18 |
| NetInv | 20.20 | 23.59 | 17.61 |

## 5. Conclusions

A manifold learning-assisted convolutional neural network, called NetInv, has been generated for structural defect reconstructions using the information of reflection coefficients obtained by solving guided wave scattering problems. Applying the hybrid finite element method, a dataset contains three representative classes of defects has been created for the purpose of training and testing the NetInv model. Throughout the numerical examples, it is demonstrated that NetInv has the ability to realize an effective mapping between the data of reflection coefficients in the wavenumber domain and defect profiles in the spatial domain, and outperforms the boundary integral equations-based reconstruction method in terms of accuracy and efficiency. Results show that NetInv has the power to achieve the defect reconstruction using the corrupted data with less than one second and produce nearly $200\%$ more accurate results than WNST. This research provides valuable insight into the development of effective data driven structural health monitoring and defect reconstruction using machine learning.